\documentclass[prd,amsmath,floats,amssymb,authoryear,floatfix,superscriptaddress,nofootinbib,onecolumn]{revtex4} 
\usepackage{graphicx}
\usepackage{hhline} 
\usepackage{caption}
\usepackage{natbib}
\defcitealias{fazzari_2025}{F25}
\usepackage{appendix}
\usepackage{multirow}
\usepackage{makecell}
\usepackage{placeins}

\renewcommand{\thetable}{\textbf{\arabic{table}}}

\makeatletter
\renewcommand{\fnum@figure}{\textbf{Figure~\thefigure}}
\renewcommand{\fnum@table}{\textbf{Table~\thetable}}
\makeatother

\usepackage{float}
\usepackage[caption=false]{subfig}
\usepackage[paperwidth=210mm,paperheight=297mm,centering,hmargin=1cm,vmargin=2.5cm]{geometry}

\usepackage[plainpages=false, colorlinks=true, anchorcolor=blue, linkcolor=blue, citecolor=blue, bookmarks=false]{hyperref}
\begin{document}
\preprint{APS/123-QED}

\title{Testing the cosmic distance duality relation using model-independent approach}

\author{Shubham Barua}
 \altaffiliation{Email:ph24resch01006@iith.ac.in}
\author{Sujit K. Dalui}
 \altaffiliation{Email:ph24mscst11036@iith.ac.in}
 \affiliation{
 Department of Physics, IIT Hyderabad Kandi, Telangana 502284,  India}
\author{Rikiya Okazaki}
 \altaffiliation{Email: r.okazaki663@gmail.com}
\affiliation{Department of Mathematical Engineering and Information Physics, The University of Tokyo, Tokyo, Japan}
\author{Shantanu Desai}
 \altaffiliation{Email:shntn05@gmail.com}
\affiliation{
 Department of Physics, IIT Hyderabad Kandi, Telangana 502284,  India}




\begin{abstract}
In this work, we test the cosmic distance duality relation (CDDR) using the arbitrary redshift pivot Pad\'e-(2,1) expansion methodology developed in Ref.~\cite{fazzari_2025}. This approach allows us to constrain the cosmography parameters and test the CDDR at any redshift. Further, it does not rely on data reconstructions or extrapolations of the cosmography parameters to higher redshifts. We employ observational data from the Dark Energy Spectroscopic Instrument (DESI) Baryon Acoustic Oscillation dataset, cosmic chronometers (CC), and Type Ia supernovae from the Pantheon Plus (PP) and Dark Energy Survey Year 5 (DESY5) compilations. We find no significant deviations from the standard CDDR relation in the range $0\lesssim z \lesssim 1$ when considering DESI$+r_d$ dataset in combination with PP$+$CC and DESY5$+$CC datasets. However, on imposing a Gaussian prior on $M_B \in \mathcal{N}(-19.253, 0.027)$ (instead of treating it as a free parameter) in the dataset combination PP$+$CC, we find CDDR violation at a level of $(3-5)\sigma.$ 
\end{abstract}


\maketitle
\section{Introduction}
\label{sec:1}

The cosmic distance duality relation (CDDR)~\cite{etherington_1933} is one of the fundamental tenets in cosmology. It is based on three fundamental assumptions~\cite{bassett_2004}: how the metric theory of gravity describes space-time, propagation of photons along null geodesics, and photon number conservation. It does not assume any information about the expansion history of the universe and hence, is agnostic to the underlying cosmological model. The CDDR establishes a relation between two distance measures for an astronomical object - the luminosity distance ($D_L(z)$) and the angular diameter distance ($D_A(z)$). Mathematically, the relation is given by $D_L(z) = (1+z)^2D_A(z)$. Due to its fundamental nature, testing this relation implies testing the foundations of modern cosmology itself~\cite{avgoustidis_2010, goncalves_2012}. Any violation of CDDR would point to a whole bunch of new physics scenarios as recently reviewed in ~\cite{zhang_2025}.

With the advent of precision cosmology, various tensions have come under scrutiny in the $\Lambda$CDM framework: the most prominent being the Hubble tension~\cite{bernal_2016,bethapudi16, vagnozzi_2020, vagnozzi_2023,  james_2022, valentino_2021, verde_2019,  abdalla_2022, scherer_2025, poulin_2024, das_2024}.
The Hubble tension refers to the discrepancy in measurements of the present expansion rate of the universe ($H_0$) between the CMB data and low-redshift probes and with statistical significance greater than $5\sigma$.
Studies indicate that any violations of CDDR  may signal inconsistencies in distance calibrations among various cosmological probes \cite{teixeira_2025, hernandez_2025}, which could alleviate the observed tensions. Hence, assessing the validity of the CDDR is crucial in the context of current cosmological discrepancies.
In light of these tensions, it has become important to consider model-independent approaches to analyze cosmological data~\cite{saikat_2021, mehrabi_2021, naik_2023, cozzumbo_2025, ormondroyd_2025, ormondroyd2_2025, berti_2025}. Various such techniques exist in the literature, such as  Artificial Neural Networks (ANN)~\cite{goh_2024, wang_2020, wang_2021,Bora22, qi_2023, abedin_2025}, cosmography~\cite{busti_2015, tucker_2005, duncsby_2016, shubham_2025}, Gaussian Process Regression (GPR) methods~\cite{Seikel,jie_2020, Colgain21, shifan_2024, Velazquez_2024, Mukherjee_2023,Haveesh} and many more~\cite{shah_2024, shah2_2024, Shah_2023, Mukherjee_2023}. Over the years, a plethora of  combinations of cosmological probes and methodologies have been utilized to test CDDR. (See ~\cite{liu_2021, li_2025, qi_2025, holanda_2010, khedekar_2011, ma_2018, holanda_2017, more_2016, zhou_2021, qi_2019, bora_2021, mukherjee_2021, xu_2022, yang_2019, alfano_2025, zheng_2025, zhang_2025, wang_2024, dasilva_2020, goncalves_2020, dacosta_2015, holanda_2012, goncalves_2012, holanda_2019} for a non-exhaustive list). In most of the works which test CDDR using cosmography~\cite{yang_2024, jesus_2025}, the constraints on the cosmography parameters are obtained at redshift $z=0$. Then, considering a parameterization for $\eta(z) = \frac{D_L(z)}{D_A(z)(1+z)^2}$, the CDDR relation is tested over a wide range of redshifts. This process has the limitation that this test depends on extrapolation of cosmography parameters. Most recently, ~\citet{fazzari_2025} (henceforth, F25) introduced generalized Pad\'e-(2,1) expansions around arbitrary pivot redshifts. This has the advantage of constraining the cosmography parameters at any fiducial redshift.

In this work, we use the Pad\'e-(2,1) expansions introduced in F25 to test CDDR at the pivot redshifts. This approach is advantageous because it does not rely on any extrapolation or reconstruction techniques or specific parameterizations. This paper is organized as follows. In Sec.~\ref{sec:2} we introduce the datasets considered and the outline our methodology, in Sec.~\ref{sec:3} we present our results and finally, in Sec.~\ref{sec:4} we summarize our findings.

\section{Methodology}
\label{sec:2}

In this section we describe the datasets employed and the methodology adopted to test the CDDR. 
\begin{itemize}
    \item \textbf{Baryon Acoustic Oscillations} (BAO): We use the recent Dark Energy Spectroscopic Instrument (DESI) data release 2~\cite{desi_2025}. The measurements we consider are $D_V/r_d$ (for BGS), $D_M/r_d$, $D_H/r_d$ and the correlations between them from different tracers for seven redshift bins centered around 0.295, 0.51, 0.71, 0.93, 1.32, 1.48, and  2.33. $D_M(z)$ is the transverse comoving distance, $D_H(z)$ is the Hubble distance and $D_V(z)$ is the angle averaged distance, all normalized to the sound horizon at the drag epoch ($r_d$). Since cosmography methods like Pad\'e and Taylor expansions are accurate only at low redshifts, they do not change early universe physics. Hence, in our analyses we consider a Planck-derived Gaussian prior on $r_d$ taken to be $\mathcal{N}(147.09, 0.26)$ Mpc~\cite{planck_2020}. 
    \item \textbf{Cosmic Chronometers} (CC): Cosmic Chronometers directly measure the Hubble expansion rate $H(z)$ by using massive, passively evolving galaxies as cosmic clocks~\cite{jimenez_2002, borghi_2022}. Motivated by the concerns raised in Ref.~\cite{kjerrgren_2023}, we restrict our analysis to a subset of 15 data points~\cite{moresco_2012, moresco_2015, moresco_2016} for which the full covariance information, including correlations and systematics, is available~\cite{moresco_2018, moresco_2020}.
    \item \textbf{Supernovae} (SNe): We use PantheonPlus (PP)~\cite{scolnic_2022} and Dark Energy Survey (DES) Y5 SNe~\cite{des_2024} samples in this work. PP consists of 1701 light curves of 1550 distinct Type Ia SNe. We remove SNe for $z\lesssim0.01$ due to strong peculiar velocity dependence~\cite{brout_2022}. DESY5 consists of 1829 Type Ia SNe (1635 in the range $0.1 < z < 1.13$ and 194 external low-redshift sample spanning $0.025<z<0.1$). When using PP, we impose both a uniform prior on the peak absolute magnitude: $M_B \in \mathcal{U}(-21, -18)$~\cite{chen_2024} and a Gaussian prior $M_B \in \mathcal{N}(-19.253, 0.027)$ based on Cepheid calibration of Type Ia SNe based on SH0ES observations~\cite{shoes_2022}. For DESY5, $M_B$ is marginalized over so no prior is required in this case.
\end{itemize}

Cosmography~\cite{busti_2015, dunsby_2016, tucker_2005} describes the expansion history of the universe in a model-independent way by expressing cosmological observables in terms of derivatives of the scale factor or the expansion rate. This is done by carrying out a Taylor expansion of the  cosmological observables about redshift $z=0$. Observational data are then used to constrain the cosmography parameters~\cite{visser_2004} at the present epoch: Hubble constant $\left(H_0 \equiv  \left.\frac{\dot{a}}{a}\right|_{z=0}\right)$, the deceleration parameter $\left(q_0 \equiv  -\left.\frac{1}{H_0^2} \frac{\ddot{a}}{a}\right|_{z=0}\right)$, the jerk parameter $\left(j_0 \equiv  \left.\frac{1}{H_0^3} \frac{\dddot{a}}{a}\right|_{z=0}\right)$ and the snap parameter $\left(s_0 = \left.\frac{1}{H_0^4}\frac{\ddddot{a}}{a}\right|_{z=0}\right)$. The limitation of Taylor expanding observables is evident. It diverges at higher redshifts $(z \gtrsim1)$, which limits its applicability given that most datasets probe redshifts in that region~\cite{aviles_2012}. 

To overcome this,  many other series  expansion strategies have been proposed~\cite{vitagliano_2010, bamba_2012}, the most popular being the Pad\'e approximants~\cite{capozziello_2020, gruber_2014,aviles_2014, wei_2014}. Following the methodology of \citetalias{fazzari_2025}, we use the Pad\'e-(2,1) expansion as it provides the best compromise between accuracy, stability and the number of kinematic parameters~\cite{capozziello_2019, hu_2022, yu_2025}. \citetalias{fazzari_2025} derived expressions for $H^{(2,1)}(z)$ and $D_L^{(2,1)}(z)$ expanded about an arbitrary pivot redshift $z_0$. This allows one to overcome the limitations of the traditional Pad\'e approximants by being able to constrain cosmography parameters at any particular redshift value directly and improve the precision of fitting cosmological data at higher redshifts. Using these expressions (presented below), we constrain the cosmography parameters over the redshift range $0 \leq z \leq 1$, with a step size of 0.1. For this purpose, we utilize different combinations of datasets (described above) specifically, BAO$+r_d$ and SNe$+$CC where SNe is taken to be either PP or DESY5. We also use the dataset PP$+$CC$+M_B$ to isolate the effect of introducing a Gaussian prior on $M_B$ instead of keeping it as a free parameter.

The cosmography expansions for $H^{(2,1)}(z)$ and $D_L^{(2,1)}(z)$ are as follows~\cite{fazzari_2025}: 
\begin{equation}
\label{eqn:1}
    f^{(2,1)}(z) = a_0+(a_1-a_0b_1)x+(a_2-a_1b_1+a_0b_1^2)x^2+(-a_2b_1+a_1b_1^2-a_0b_1^3)x^3 + \mathcal{O}(x^4),
\end{equation}
where the Pad\'e parameters $(a_0, a_1, a_2, b_1)$ can be fixed by comparing to the Taylor expansions of $H(z)$ and $D_L(z)$ which are given by 
\begin{align}
    H^\text{Tay}(z) &= c_0+c_1x+c_2x^2+c_3x^3+\mathcal{O}(x^4) \label{eqn:2}, \text{where}\\
    c_0 &=H(z_0) \label{eqn:3}\\
    c_1 &=H(z_0)A \label{eqn:4}\\
    c_2 &=H(z_0)B \label{eqn:5}\\
    c_3 &=H(z_0)C  \label{eqn:6} \text{ and}\\
    D_L^\text{Tay}(z) &= c_0+c_1x+c_2x^2+c_3x^3+\mathcal{O}(x^4) \label{eqn:7}, \text{where}\\
    c_0 &=\frac{1+z_0}{H(z_0)}\left[z_0+\frac{A}{2}z_0^2+\frac{A^2-B}{3}z_0^3\right] \label{eqn:8}\\
    c_1 &=\frac{1}{H(z_0)}\left[1+2z_0+\frac{A}{2}z_0^2+\frac{A^2-B}{3}z_0^3\right] \label{eqn:9}\\
    c_2 &=\frac{1}{2H(z_0)}\left[2-A(1+z_0)\right] \label{eqn:10}\\
    c_3 &=\frac{1}{6H(z_0)}\left[2(1+z_0)(A^2-B)-3A\right]. \label{eqn:11}
\end{align}
In Eqns.~\ref{eqn:2}-\ref{eqn:11}, $A = \frac{1+q(z_0)}{1+z_0}$, $B = \frac{j(z_0)-q(z_0)^2}{2(1+z_0)^2}$ and $C = \frac{3q(z_0)^3+3q(z_0)^2-(4q(z_0)+3)j(z_0) - s(z_0)}{6(1+z_0)^3}$. Then, we fix the Pad\'e parameters in Eqn.~\ref{eqn:1} using $a_0=c_0$, $a_1=c_1+c_0b_1$, $a_2=c_2+c_1b_1$ and $b_1=-\frac{c_3}{c_2}$. Finally, using these Pad\'e parameters, we can get the Pad\'e-(2,1) expressions for $H^{(2,1)}(z)$ and $D_L^{(2,1)}(z)$ using
\begin{equation}
\label{eqn:12}
    f^{(2,1)}(z) = \frac{a_0 + a_1x + a_2x^2}{1 + b_1x}
\end{equation}

To test CDDR, we use the relation 
\begin{equation}
\label{eqn:13}
    \eta(z=z_0) = \frac{D_L(z_0)}{(1+z_0)^2D_A(z_0)},
\end{equation}
where, $D_L(z)$ is the luminosity distance at redshift $z$ and $D_A(z)$ is the angular diameter distance, given by
\begin{align}
    D_L(z) &= c(1+z)\int_0^z\frac{dz'}{H(z')}  \label{eqn:14} \\
    D_A(z) &= \frac{D_M(z)}{1+z}. \label{eqn:15}
\end{align}
$D_M(z)$, $D_V(z)$, and $D_H(z)$ are given by
\begin{align}
    D_M(z) &= c\int_0^z\frac{dz'}{H(z')} \label{eqn:16} \\
    D_H(z) & = \frac{c}{H(z)} \label{eqn:17} \\
    D_V(z) &= [zD_M(z)^2D_H(z)]^{1/3}. \label{eqn:18}
\end{align}
Finally, the relation between $M_B$ and $D_L(z)$ for Type Ia SNe is given by:
\begin{equation}
\label{eqn:19}
    m(z) = 5\log_{10}\left[\frac{D_L(z)}{\text{Mpc}}\right] + 25 + M_B,
\end{equation}
where $m(z)$ denotes the SNe apparent magnitude.

We use \texttt{Cobaya} sampler~\cite{cobaya} to perform Markov Chain Monte Carlo (MCMC) sampling~\cite{lewis_2002, lewis_2013} in order to constrain the cosmography parameters at the pivot redshift $z_0$. We consider the chains to have converged for the Gelman-Rubin criterion of $R-1\lesssim0.01$ \cite{gelman_1992}. We use \texttt{GetDist}~\cite{getdist_2025} for analysis and visualization of the Bayesian posteriors. In this work, we consider a flat universe (Eqn.~\ref{eqn:16}).

To test CDDR, we first compute $D_A(z_0)$ by inserting the cosmography parameters constrained from the BAO$+r_d$ dataset in Eqn.~\ref{eqn:15}. Similarly, we compute $D_L(z_0)$ by using the constraints on the cosmography parameters derived from the SNe$+$CC and PP$+$CC$+M_B$ datasets and evaluating Eqn.~\ref{eqn:14}. Finally, we form the ratio in Eqn.~\ref{eqn:13} to obtain $\eta(z_0)$. For each MCMC run, we extract the full posterior samples of the cosmography parameters. 

Since the error bars on $\eta(z_0)$ are asymmetric, we symmetrize using the following equation:~\cite{cao_2022}
\begin{equation}
\label{eq29}
    \sigma_s = \frac{1}{2}\left[2\frac{\sigma_{+}\sigma_{-}}{\sigma_{+}+\sigma_{-}}+\sqrt{\sigma_{+}\sigma_{-}}\right],
\end{equation}
where $\sigma_s$ is the symmetrized uncertainty, and $+$ and $-$ denote the upper and lower error bars, respectively.

\section{Results and Discussion}
\label{sec:3}

\begin{figure}
    \centering
    \includegraphics[width=1.0\textwidth]{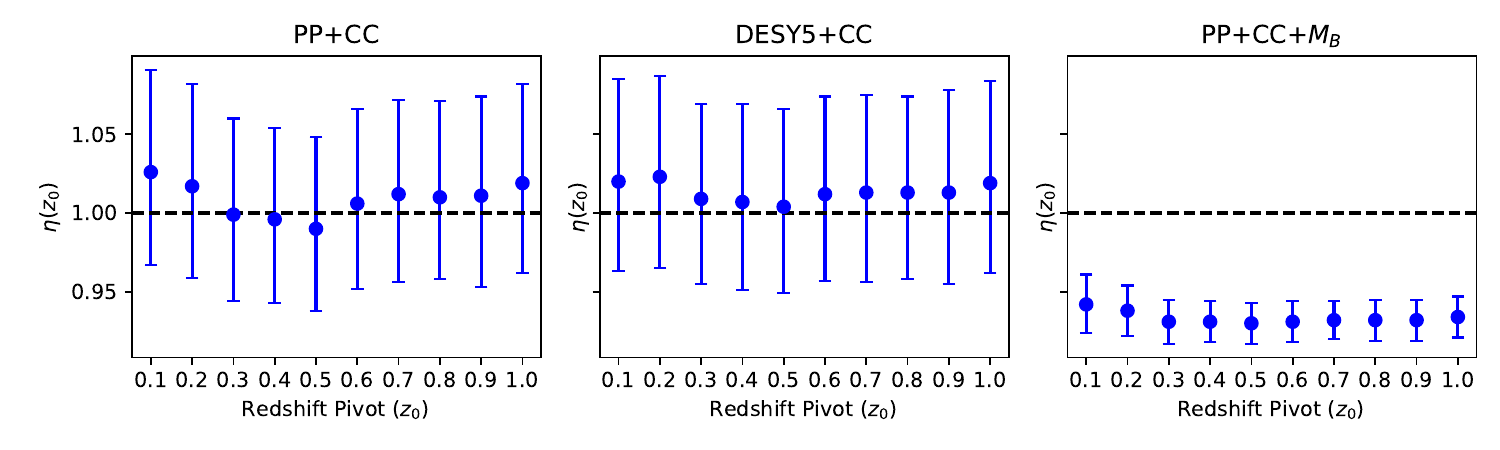}
    \caption{Constraints on $\eta(z_0)$ along with 68\% credible intervals for SNe$+$CC dataset (where SNe is either PP or DESY5) in combination with BAO$+r_d$ at the pivot redshifts $z_0\in[0,1]$.}
    \label{fig1}
\end{figure}

The values of the cosmography parameters are listed in Table~\ref{table1}. The deviations of $\eta(z_0)$ from its standard value of 1 are reported in the last column. Figure~\ref{fig1} provides a visual representation of the constraints on $\eta(z_0)$ for each dataset combination. 

For DESI$+r_d$, PP$+$CC and DESY5$+$CC we recover identical constraints on the cosmography parameters and the same overall trends as reported in \citetalias{fazzari_2025}.

Since we use BAO$+r_d$ dataset to constrain $D_A(z_0)$ and SNe$+$CC dataset to constrain $D_L(z_0)$, there is no issue of correlated uncertainties as the two dataset combinations are statistically independent. Moreover, we account for all uncertainties because we propagate the entire posterior distributions of the cosmography parameters from each MCMC chain when computing $\eta(z_0)$. With this we can confirm that the check for CDDR is cosmology-independent.

Our results indicate no violation of CDDR within $1\sigma$ for SNe$+$CC dataset combined with BAO$+r_d$ dataset. The uncertainties associated with $\eta(z_0)$ lie at the percent-level, typically in the range $(5-7)\%$.

Using the PP$+$CC$+M_B$ dataset to constrain $D_L(z_0)$ we see that there is a deviation of $\sim (3-5)\sigma$ of CDDR parameter $\eta(z_0)$ from its theoretical value of 1. This is due to the tension that exists between the early universe inferred $r_d$ value and the late universe obtained $M_B$ value.

We also note  that the constraints on the cosmography parameters obtained using PP$+$CC$+M_B$ dataset are similar to the one found using PP$+$CC. Only the $H_0$ values from PP$+$CC$+M_B$ dataset are higher than those using found using PP$+$CC dataset since the $M_B$ calibration pushes $H_0$ to be higher. Further the error bars on $H_0$ are comparatively smaller than in PP$+$CC due to the breaking of the degeneracy between $H_0$ and $M_B$.

\begin{table}
\caption{Cosmography parameter constraints at 68\% credible intervals for pivot redshifts in the range $z_0\in[0,1]$. The constraints on the CDDR parameter $\eta(z=z_0)$ and its deviation from $\eta(z_0) = 1$ is also shown. For each pivot redshift, cosmography constraints are shown for all datasets. As described in the text, BAO$+r_d$ constrains $D_A(z_0)$ while SNe$+$CC constrains $D_L(z_0)$. Hence, $\eta(z_0)$ and its deviations are reported only for the SNe$+$CC combinations. It is understood that the SNe$+$CC and BAO$+r_d$ datasets are combined as specified above.}
\centering
\begin{tabular}{c l l l l l l l}

\hline
\thead{Redshift} & \thead{Dataset} & \thead{$H(z_0)$} & \thead{$q(z_0)$} & \thead{$j(z_0)$} & \thead{$s(z_0)$} & \thead{$\eta(z_0)$} & \thead{Deviations($\sigma$)} \\
 & & [km/s/Mpc] & & & & & \\
\hline

    \noalign{\vskip 4pt}
    $z_0=0.0$ & DESI$+r_d$            & $68.4 \pm 1.5$ & $-0.53 \pm 0.16$&$1.12_{-0.79}^{+0.36}$ &$0.61^{+0.19}_{-1.2}$ & &  \\
              & PP$+$CC               & $67.3 \pm 4.4$ & $-0.51 \pm 0.07$&$1.38 \pm 0.54$ &$1.8^{+1.6}_{-2.2}$ & - & - \\
              & PP$+$CC$+M_B$         & $72.94 \pm 0.9$ & $-0.50^{+0.06}_{-0.07}$&$1.24 \pm 0.50$ &$>1.41$ & - & - \\
              & DESY5$+$CC            & $67 \pm 4.3$ & $-0.43^{+0.07}_{-0.10}$&$0.97^{+0.66}_{-0.55}$ &$1.2^{+2.4}_{-1.9}$ & - & - \\
    \noalign{\vskip 4pt} 
    \hline
    \noalign{\vskip 4pt}
    $z_0=0.1$ & DESI$+r_d$            & $71.87 \pm 0.89$ & $-0.42^{+0.13}_{-0.12}$&$1.03_{-0.72}^{+0.33}$ &$0.13^{+2.0}_{-0.7}$ & & \\
              & PP$+$CC               & $70.6 \pm 4.3$ & $-0.41 \pm 0.04$&$1.27 \pm 0.44$ &$1.4^{+1.7}_{-2.5}$ & $1.026^{+0.065}_{-0.059}$ & $0.42$\\
              & PP$+$CC$+M_B$         & $76.93 \pm 0.97$ & $-0.4 \pm 0.04$&$1.13 \pm 0.46$ &$1.8^{+2.6}_{-1.7}$ & $0.942^{+0.019}_{-0.018}$ & $3.13$ \\
              & DESY5$+$CC            & $71.5 \pm 4.5$ & $-0.34 \pm 0.05$&$0.78^{+0.50}_{-0.57}$ &$0.4^{+2.4}_{-1.8}$ & $1.017^{+0.070}_{-0.061}$ & $0.26$ \\
    \noalign{\vskip 4pt} 
    \hline
    \noalign{\vskip 4pt}
    $z_0=0.2$ & DESI$+r_d$            &  $75.88 \pm 0.46$ & $-0.32^{+0.10}_{-0.08}$ & $0.95^{+0.28}_{-0.64}$ &$-0.43^{+1.74}_{-0.48}$ &  &  \\
              & PP$+$CC                & $74.8 \pm 4.6$ & $-0.33 \pm 0.03$&$1.42 \pm 0.40$ &$1.0^{+2.0}_{-2.6}$ & $1.017^{+0.065}_{-0.058}$ & $0.28$ \\
              & PP$+$CC$+M_B$         & $81.2 \pm 1.1$ & $-0.33 \pm 0.03$&$1.33 \pm 0.39$ &$>0.52$ & $0.938\pm0.016$ & $3.88$ \\
              & DESY5$+$CC            & $74.9 \pm 4.5$ & $-0.29 \pm 0.03$&$1.17 \pm 0.53$ &$1.1^{+2.2}_{-2.6}$ & $1.021^{+0.065}_{-0.058}$ & $0.34$ \\
    \noalign{\vskip 4pt} 
    \hline
    \noalign{\vskip 4pt}
    $z_0=0.3$ & DESI$+r_d$            &  $80.44^{+0.57}_{-0.44}$ & $-0.24^{+0.058}_{-0.048}$&$0.85^{+0.21}_{-0.50}$ &$-0.97^{+1.04}_{-0.22}$ &  &  \\
              & PP$+$CC                & $80.7 \pm 4.8$ & $-0.26 \pm 0.03$&$0.68  \pm 0.13$ &$0.99^{+1.6}_{-2.9}$ & $0.999^{+0.061}_{-0.055}$ & $0.017$ \\
              & PP$+$CC$+M_B$         & $86.6 \pm 1.1$ & $-0.26 \pm 0.03$&$0.68 \pm 0.13$ &$1.2^{+1.7}_{-2.7}$ & $0.931\pm0.014$ & $4.93$ \\
              & DESY5$+$CC            & $79.7 \pm 4.8$ & $-0.24 \pm 0.03$&$1.09^{+0.33}_{-0.25}$ &$0.6^{+2.0}_{-2.9}$ & $1.010^{+0.065}_{-0.056}$ & $0.17$ \\
    \noalign{\vskip 4pt} 
    \hline
    \noalign{\vskip 4pt}
    $z_0=0.4$ & DESI$+r_d$            &  $85.41^{+0.76}_{-0.53}$ & $-0.16 \pm 0.03$&$0.80^{+0.22}_{-0.41}$ &$-1.31^{+0.69}_{-0.17}$ &  &  \\
              & PP$+$CC               & $85.6 \pm 4.9$ & $-0.20 \pm 0.03$&$0.56 \pm 0.05$ &$0.9^{+2.1}_{-2.7}$ & $0.996^{+0.058}
              _{-0.053}$ & $0.072$ \\
              & PP$+$CC$+M_B$         & $91.6 \pm 1.2$ & $-0.21 \pm 0.03$&$0.56 \pm 0.05$ &$1.0^{+1.8}_{-2.7}$ &$0.931\pm0.013$ & $5.31$ \\
              & DESY5$+$CC            & $85.2^{+4.6}_{-5.2}$ & $-0.18 \pm 0.03$&$0.68 \pm 0.10$ &$0.7^{+1.9}_{-2.9}$ & $1.005^{+0.059}_{-0.056}$ & $0.09$ \\
    \noalign{\vskip 4pt} 
    \hline
    \noalign{\vskip 4pt}
    $z_0=0.5$ & DESI$+r_d$            &  $90.82^{+0.73}_{-0.65}$ & $-0.11 \pm 0.03$&$0.73^{+0.23}_{-0.29}$ &$-1.52^{+0.10}_{-0.33}$ &  &  \\
              & PP$+$CC               & $90.4 \pm 5.1$ & $-0.15 \pm 0.04$&$0.57^{+0.03}_{-0.04}$ &$0.4 \pm 2.4$ & $0.99^{+0.058}_{-0.052}$ & $0.18$ \\
              & PP$+$CC$+M_B$         & $97.0 \pm 1.5$ & $-0.16 \pm 0.04$&$0.57 \pm 0.03$ &$0.3^{+2.1}_{-2.4}$ & $0.93\pm0.013$ & $5.38$ \\
              & DESY5$+$CC            & $90.1 \pm 5.3$ & $-0.12 \pm 0.04$&$0.64 \pm 0.04$ &$0.6^{+2.3}_{-2.8}$ & $1.008^{+0.060}_{-0.055}$ & $0.14$ \\
    \noalign{\vskip 4pt} 
    \hline
    \noalign{\vskip 4pt}
    $z_0=0.6$ & DESI$+r_d$            &  $96.43^{+0.81}_{-0.72}$ & $-0.05^{+0.02}_{-0.03}$&$0.74^{+0.21}_{-0.26}$ &$-1.76^{+0.13}_{-0.38}$ &  &  \\
              & PP$+$CC               & $95.3 \pm 5.6$ & $-0.09 \pm 0.04$&$0.62^{+0.04}_{-0.05}$ &$-0.2^{+2.1}_{-2.4}$ & $1.006^{+0.060}_{-0.054}$ & $0.11$ \\
              & PP$+$CC$+M_B$         & $102.8 \pm 1.8$ & $-0.1 \pm 0.04$&$0.61^{+0.04}_{-0.05}$ &$-0.4^{+1.9}_{-2.1}$ & $0.931\pm0.013$ & $5.31$ \\
              & DESY5$+$CC            & $95.6 \pm 5.6$ & $-0.07 \pm 0.04$&$0.67^{+0.04}_{-0.05}$ &$0.0^{+2.0}_{-2.6}$ & $1.007^{+0.062}_{-0.055}$ & $0.12$ \\ 
    \hline
    \noalign{\vskip 4pt} 

    $z_0=0.7$ & DESI$+r_d$            &  $102.38 \pm 0.7$ & $-0.01^{+0.02}_{-0.03}$&$0.73^{+0.18}_{-0.20}$ &$-2.10^{+0.19}_{-0.34}$ &  &  \\
              & PP$+$CC               & $100.6 \pm 5.9$ & $-0.03 \pm 0.04$&$0.69\pm0.06$ &$-0.8^{+2.0}_{-2.3}$ & $1.012^{+0.060}_{-0.056}$ & $0.21$ \\
              & PP$+$CC$+M_B$         & $109.0 \pm 2.0$ & $-0.04^{+0.04}_{-0.05}$&$0.68^{+0.05}_{-0.07}$ &$-1.2^{+1.6}_{-1.9}$ & $0.932\pm0.012$ & $5.67$ \\
              & DESY5$+$CC            & $101.2 \pm 5.9$ & $-0.01 \pm 0.04$&$0.73^{+0.05}_{-0.06}$ &$-0.8 \pm 1.9$ & $1.009^{+0.060}_{-0.055}$ & $0.16$ \\
    \noalign{\vskip 4pt} 
    \hline
    \noalign{\vskip 4pt}
    $z_0=0.8$ & DESI$+r_d$            &  $108.51 \pm 0.74$ & $0.03^{+0.03}_{-0.04}$&$0.75^{+0.20}_{-0.17}$ &$-2.37^{+0.24}_{-0.44}$ &  &  \\
              & PP$+$CC               & $107.1 \pm 6.3$ & $0.04^{+0.05}_{-0.04}$&$0.77 \pm 0.07$ &$-1.6 \pm 1.8$ & $1.010^{+0.061}_{-0.052}$ & $0.18$ \\
              & PP$+$CC$+M_B$         & $115.8 \pm 2.4$ & $0.02^{+0.05}_{-0.04}$&$0.75 \pm 0.07$ &$-2.1^{+1.4}_{-2.1}$ & $0.932\pm0.013$ & $5.23$ \\
              & DESY5$+$CC            & $107.6 \pm 6.3$ & $0.05 \pm 0.05$&$0.81^{+0.07}_{-0.08}$ &$-1.6^{+1.8}_{-2.1}$ & $1.009^{+0.061}_{-0.055}$ & $0.16$ \\
    \noalign{\vskip 4pt} 
    \hline
    \noalign{\vskip 4pt}
    $z_0=0.9$ & DESI$+r_d$            &  $114.81 \pm 0.74$ & $0.06^{+0.03}_{-0.04}$&$0.78^{+0.17}_{-0.13}$ &$-2.78^{+0.28}_{-0.46}$ &  &  \\
              & PP$+$CC               & $114.5 \pm 7.0$ & $0.13\pm0.06$&$0.91^{+0.09}_{-0.13}$ &$-2.1 \pm 1.6$ & $1.011^{+0.063}_{-0.058}$ & $0.18$ \\
              & PP$+$CC$+M_B$         & $123.5^{+2.8}_{-3.2}$ & $0.11^{+0.05}_{-0.06}$&$0.87^{+0.08}_{-0.11}$ &$-2.6^{+1.0}_{-2.1}$ & $0.932\pm0.013$ & $5.23$ \\
              & DESY5$+$CC            & $114.2 \pm 7.1$ & $0.13^{+0.05}_{-0.06}$&$0.93^{+0.09}_{-0.12}$ &$-2.1 \pm 1.6$ & $1.015^{+0.065}_{-0.058}$ & $0.24$ \\
    \noalign{\vskip 4pt} 
    \hline
    \noalign{\vskip 4pt}
    $z_0=1.0$ & DESI$+r_d$            &  $121.34 \pm 0.8$ & $0.09\pm0.04$&$0.82^{+0.16}_{-0.11}$ &$-3.2^{+0.36}_{-0.58}$ &  &  \\
              & PP$+$CC               & $121.6 \pm 7.2$ & $0.22^{+0.07}_{-0.08}$&$1.09^{+0.12}_{-0.18}$ &$-2.6 \pm 1.3$ & $1.019^{+0.063}_{-0.057}$ & $0.32$ \\
              & PP$+$CC$+M_B$         & $131.4 \pm 3.5$ & $0.18 \pm 0.06$&$1.0 \pm 0.11$ &$<-2.57$ & $0.934\pm0.013$ & $5.07$ \\
              & DESY5$+$CC            & $121.3\pm 7.3$ & $0.21^{+0.07}_{-0.06}$&$1.07 \pm 0.13$ &$-2.6^{+1.3}_{-1.6}$ & $1.019^{+0.066}_{-0.056}$ & $0.31$\\
    \noalign{\vskip 4pt} 
    \hline
\label{table1}

\end{tabular}
\end{table}

\section{Conclusions}
\label{sec:4}

In this work, we test the cosmic distance duality relation (CDDR) by applying the methodology of \citetalias{fazzari_2025} to a combination of CC, DESI BAO, and SNe data from the PP and DESY5 compilations. Specifically, we employ the generalized Pad\'e-(2,1) expansion about arbitrary pivot redshifts to obtain model-independent constraints on the CDDR parameter $\eta(z)$ (cf. Eqn.~\ref{eqn:13}). This approach allows us to avoid extrapolating cosmography parameters constrained at redshift $z=0$ to higher redshifts. Moreover, our method does not require dataset reconstructions, which can introduce additional uncertainties, noise, or dependence on binning choices. Instead, we  determine the values of the cosmography parameters and $\eta(z)$ at a chosen pivot redshift. In this sense, our analysis represents the first test of the CDDR performed directly at specific redshifts.

We find no CDDR violation within $1\sigma$ in the redshift range $[0,1]$ for the dataset SNe$+$CC. However, on imposing a Gaussian prior on $M_B$ (instead of treating it as a free parameter), we notice CDDR violation at the $\sim(3-5)\sigma$ level. This is expected since there exists a tension between the value of $r_d$ inferred from early universe observations and the $M_B$ value obtained from late universe data~\cite{shubham_2025, poulin_2024}. This CDDR violation was also noticed in~\cite{li_2025}. Hence, calibration plays an important role in determining whether CDDR is violated or not. It will be interesting to see what happens at higher redshifts when more data is available.

Overall, this work demonstrates the application of the arbitrary-pivot Pad\'e-(2,1) expansion introduced in \citetalias{fazzari_2025}. We also conclude that our results show no evidence for a violation of the CDDR when using uncalibrated SNe dataset.

\begin{acknowledgments}
SB would like to extend his gratitude to the University Grants Commission (UGC), Govt. of India for their continuous support through the Junior Research Fellowship, which has played a crucial role in the successful completion of our research. RO would like to thank the JST Sakura Science Program (Grant No. M2025L0418003) and the JASSO Scholarship for supporting his travel to and stay at IIT Hyderabad.
Computational work was supported by the National Supercomputing Mission (NSM), Government of India, through access to the ``PARAM SEVA'' facility at IIT Hyderabad. The NSM is implemented by the Centre for Development of Advanced Computing (C-DAC) with funding from the Ministry of Electronics and Information Technology (MeitY) and the Department of Science and Technology (DST). We would like to thank Eoin \'O Colg\'ain for valuable comments on this work. We are also grateful to the anonymous referee for useful feedback and comments on the manuscript.
\end{acknowledgments}

\clearpage
\bibliography{references}

\end{document}